\input harvmac
\noblackbox

\font\cmss=cmss10 \font\cmsss=cmss10 at 7pt
 \def\inbar{\,\vrule height1.5ex width.4pt depth0pt}
\def\IZ{\relax\ifmmode\mathchoice
{\hbox{\cmss Z\kern-.4em Z}}{\hbox{\cmss Z\kern-.4em Z}}
{\lower.9pt\hbox{\cmsss Z\kern-.4em Z}}
{\lower1.2pt\hbox{\cmsss Z\kern-.4em Z}}\else{\cmss Z\kern-.4em
Z}\fi}
\def\IB{\relax{\rm I\kern-.18em B}}
\def\IC{{\relax\hbox{$\inbar\kern-.3em{\rm C}$}}}
\def\ID{\relax{\rm I\kern-.18em D}}
\def\IE{\relax{\rm I\kern-.18em E}}
\def\IF{\relax{\rm I\kern-.18em F}}
\def\IG{\relax\hbox{$\inbar\kern-.3em{\rm G}$}}
\def\IGa{\relax\hbox{${\rm I}\kern-.18em\Gamma$}}
\def\IH{\relax{\rm I\kern-.18em H}}
\def\II{\relax{\rm I\kern-.18em I}}
\def\IK{\relax{\rm I\kern-.18em K}}
\def\IC{\relax{\rm I\kern-.18em C}}
\def\IR{\relax{\rm I\kern-.18em R}}


\lref\gibbons{G.~W.~Gibbons,
``Vacuum Polarization And The Spontaneous Loss Of Charge By Black Holes,''
Commun.\ Math.\ Phys.\  {\bf 44}, 245 (1975).}
\lref\RS{L. Randall and R. Sundrum, ``An Alternative to Compactification''
Phys.\ Rev.\ Lett.\  {\bf 83}, 4690 (1999),
hep-th/9906064
}
\lref\itzhaki{N. Itzhaki, J. Maldacena, J. Sonnenschein, and
Shimon Yankielowicz, ``Supergravity and the Large N Limit of
Theories with Sixteen Supercharges'', hep-th/9802042.}
\lref\kut{D. Kutasov,``Orbifolds and Solitons,''
Phys.\ Lett.\  {\bf B383}, 48 (1996),
hep-th/9512145. }
\lref\sen{A. Sen,``Duality and Orbifolds,''
Nucl.\ Phys.\  {\bf B474}, 361 (1996), hep-th/9604070. }
\lref\smeared{See for example P.~Kraus, F.~Larsen and S.~P.~Trivedi,
``The Coulomb branch of gauge theory from rotating branes,''
JHEP {\bf 9903}, 003 (1999),
hep-th/9811120;
M.~Cvetic, S.~S.~Gubser, H.~Lu and C.~N.~Pope,
``Symmetric
potentials of gauged supergravities in diverse
dimensions and  Coulomb branch of gauge theories,''
hep-th/9909121.
}
\lref\stringglob{T. Banks, L. Dixon,
``Constraints On String Vacua With Space-Time Supersymmetry,''
Nucl.\ Phys.\  {\bf B307}, 93 (1988).
}
\lref\conifold{A. Strominger,
``Massless black holes and conifolds in string theory,''
Nucl.\ Phys.\  {\bf B451}, 96 (1995),
hep-th/9504090.
}
\lref\wittbound{E. Witten, ``Bound States of Strings and p-branes''
Nucl.\ Phys.\  {\bf B460}, 335 (1996),
hep-th/9510135.
}
\lref\singsolutions{
O.~DeWolfe, D.~Z.~Freedman, S.~S.~Gubser and A.~Karch,
``Modeling the fifth dimension with scalars and gravity,''
hep-th/9909134.
}
\lref\oferetal{O. Aharony, M. Berkooz, D. Kutasov, and
N. Seiberg, ``Linear Dilatons, NS5-branes, and Holography'',
hep-th/9808149.}
\lref\phases{E. Witten, ``Phases of N=2 Theories in Two Dimensions''
Nucl.\ Phys.\  {\bf B403}, 159 (1993),
hep-th/9301042.
}
\lref\hermann{H. Verlinde,
``Holography and compactification,''
hep-th/9906182.
}
\lref\eddiverse{E. Witten,
``String theory dynamics in various dimensions,''
Nucl.\ Phys.\  {\bf B443}, 85 (1995),
hep-th/9503124.
}
\lref\susswitt{L. Susskind and E. Witten,
``The holographic bound in anti-de Sitter space,''
hep-th/9805114.
}
\lref\gaugebulk{
A.~Pomarol,
``Gauge bosons in a five-dimensional theory with localized gravity,''
hep-ph/9911294;
;
H.~Davoudiasl, J.~L.~Hewett and T.~G.~Rizzo,
``Bulk gauge fields in the Randall-Sundrum model,''
Phys.\ Lett.\  {\bf B473}, 43 (2000),
hep-ph/9911262.
}
\lref\RSII{L. Randall and R. Sundrum,
``A large mass hierarchy from a small extra dimension,''
Phys.\ Rev.\ Lett.\  {\bf 83} (1999) 3370
hep-ph/9905221.
}
\lref\minsei{S. Minwalla and N. Seiberg,
``Comments on the IIA NS5-brane,''
JHEP {\bf 9906}, 007 (1999),
hep-th/9904142.
}
\lref\RSBH{
A.~Chamblin, S.~W.~Hawking and H.~S.~Reall,
``Brane-world black holes,''
Phys.\ Rev.\  {\bf D61}, 065007 (2000),
hep-th/9909205;
R.~Emparan, G.~T.~Horowitz and R.~C.~Myers,
``Exact description of black holes on branes,''
JHEP {\bf 0001}, 007 (2000),
hep-th/9911043;
S.~B.~Giddings, E.~Katz and L.~Randall,
``Linearized gravity in brane backgrounds,''
JHEP {\bf 0003}, 023 (2000),
hep-th/0002091.
}
\lref\tomofer{O. Aharony and T. Banks,
``Note on the quantum mechanics of M theory,''
JHEP {\bf 9903}, 016 (1999),
hep-th/9812237.
}
\lref\glob{M.~Kamionkowski and J.~March-Russell,
``Are textures natural?,''
Phys.\ Rev.\ Lett.\  {\bf 69}, 1485 (1992),
hep-th/9201063;
R.~Holman, S.~D.~Hsu, E.~W.~Kolb, R.~Watkins and L.~M.~Widrow,
``Cosmological texture is incompatible with Planck scale physics,''
Phys.\ Rev.\ Lett.\  {\bf 69}, 1489 (1992).}
\lref\adscft{J.~Maldacena,
``The large N limit of superconformal field theories and supergravity,''
Adv.\ Theor.\ Math.\ Phys.\  {\bf 2}, 231 (1998),
hep-th/9711200;
E.~Witten,
``Anti-de Sitter space and holography,''
Adv.\ Theor.\ Math.\ Phys.\  {\bf 2}, 253 (1998)
hep-th/9802150;
S.~S.~Gubser, I.~R.~Klebanov and A.~M.~Polyakov,
``Gauge theory correlators from non-critical string theory,''
Phys.\ Lett.\  {\bf B428}, 105 (1998),
hep-th/9802109.}
\lref\duff{M.~J.~Duff and J.~T.~Liu,
``On the equivalence of the Maldacena and Randall-Sundrum pictures,''
hep-th/0003237.}
\lref\gub{S.~S.~Gubser,
``AdS/CFT and gravity,''
hep-th/9912001.}
\lref\addk{N.~Arkani-Hamed, S.~Dimopoulos, G.~Dvali and N.~Kaloper,
``Infinitely large new dimensions,''
Phys.\ Rev.\ Lett.\ {\bf 84}, 586 (2000),
hep-th/9907209.
}
\lref\rcm{R.~C.~Myers,
``Dielectric branes,''
JHEP {\bf 9912}, 022 (1999),
hep-th/9910053.
}
\lref\fivebrane{A.~Strominger,
``Heterotic Solitons,''
Nucl.\ Phys.\  {\bf B343}, 167 (1990);
C.~G.~Callan, J.~A.~Harvey and A.~Strominger,
``Supersymmetric string solitons,''
hep-th/9112030.}

\Title{\vbox{\baselineskip12pt\hbox{hep-th/0006192}
\hbox{SLAC-PUB-8482}
\hbox{SU-ITP-00/17}
}}
{\vbox{\centerline{Gauge Symmetry and Localized Gravity}\smallskip
\centerline{in M Theory}}
}

\centerline{ Nemanja Kaloper$^1$,
Eva Silverstein$^{1,2}$, and Leonard Susskind$^1$}
\bigskip
\bigskip
\centerline{$^{1}$Department of Physics}
\centerline{Stanford University}
\centerline{Stanford, CA 94305}
\medskip
\medskip
\medskip
\centerline{$^{2}$ Stanford Linear Accelerator Center}
\centerline{Stanford University}
\centerline{Stanford, CA 94309}
\bigskip
\medskip
\noindent

We discuss the possibility of having gravity ``localized'' in
dimension d in a system where gauge bosons propagate in dimension d+1.
In such a circumstance---depending on the rate of falloff of
the field strengths in d dimensions---one might expect the
gauge symmetry in d+1 dimensions to
behave like a global symmetry in d dimensions,
despite the presence of gravity.
Naive extrapolation of warped long-wavelength solutions of general
relativity coupled to scalars and gauge fields suggests
that such an effect might be possible.  However, in some basic
realizations of such solutions in M theory, we find that
this effect does not persist microscopically.  It
turns over either to screening or the Higgs mechanism at long
distances in the d-dimensional description of the system.
We briefly discuss the physics of charged objects in this
type of system.

\Date{June 2000}

\newsec{Introduction and Summary}

In the study of string dualities and the relation of string theory to
field theory, the localization of gauge dynamics
(or more general quantum field theory dynamics) to a submanifold
of spacetime has been analyzed in many contexts.  More recently the possibility
of localizing gravity has emerged in the study of cut-off $AdS$ spaces
\RS.  It is natural to wonder therefore whether it is possible
to localize gravity along a $d$-dimensional slice of spacetime
in a system where the gauge fields of some symmetry
group $G$ propagate in $d+1$ dimensions.

In the most extreme imaginable versions of
such a situation, in a $d$-dimensional description the
symmetry $G$ would appear more like a global symmetry than
a gauge symmetry.  Field lines would fall off faster than
appropriate for a
gauge symmetry in $d$-dimensions.  On the
other hand the conservation of the associated charge would
be guaranteed by the fact that the symmetry was gauged in
$d+1$ dimensions.  Because gravity is localized to $d$ dimensions,
one would then have a global symmetry in
the presence of gravity.
Since the no-hair theorems for black holes
in ordinary $d$-dimensional effective field theory suggest strongly
that such an effect is impossible \glob, it is likely that this situation
could only occur, if at all, in systems
with an infinite number of degrees of freedom in a $d$-dimensional
description.  In systems with a holographic description in terms
of a low-energy worldvolume quantum field
theory on ($d$-1)-branes, the physics must
have a conventional interpretation in $d$ dimensions.

In this paper we will obtain results consistent
with this expectation by analyzing various systems in M theory.
At the level of low-energy effective field theory
in $d+1$ dimensions, the effect appears possible,
even generic.
Consider a $d+1$-dimensional system involving a gauge theory with
field strength $F$
coupled to a scalar $\phi$ and gravity.  The action takes the form
\eqn\genac{S=\int d^dx dr\sqrt{g}\biggl(a(\phi)R+b(\phi)(\nabla\phi)^2
+c(\phi)F^2-\Lambda(\phi)\biggr).}
Here $a,b,c$ and $\Lambda$ are general functions of $\phi$.
Many such systems have ``warped'' solutions which
give a localized graviton in $d$ dimensions upon integrating
over the $d+1$st coordinate $r$ in \genac.  The
dimensional reduction of the $F^2$ term can
in general
behave differently from that of the Einstein term,
since its kinetic term involves an extra
power of the inverse metric $g^{\mu\nu}$ relative to that of
gravity, and since its coupling to the scalar will in general
differ.  At the level of \genac\ (without worrying
about its embedding into quantum gravity) there will be a large
class of systems for which the coefficient of the $F^2$ term
will
diverge.  This indicates that the long-distance
behavior of the $F$ field is weaker than that of an ordinary
unscreened un-Higgsed gauge field in $d$ dimensions.

One such case is the cutoff $AdS_5$ system with $d=4$,
which we study in \S2.\foot{Gauge fields in the bulk of
the Randall-Sundrum approach to the hierarchy problem
\RSII\ were studied by \gaugebulk.}
In this system, however, the divergence in the $F^2$ term
is logarithmic in the energy scale of the dual conformal
field theory coupled to gravity, so that the effect is identical
to that of screening of the charge in $4$ dimensions.\foot{We thank
E. Witten for pointing this out.}  We calculate the electrostatic
potential for a test charge at the ``Planck brane''
and find a result consistent with this interpretation.  In
realizations of this system in type IIB supergravity, the charged matter that
leads to the screening
is evident, although the effect
arises from
geometry independent of assuming
the presence of charged matter in the bulk.

We also study the pattern of infrared divergences in kinetic
terms for arbitrary $q$-form gauge potentials in arbitrary
dimension.  This suggests a ``screening'' phenomenon
for higher-form fields in a range of dimensions.

A more interesting case is a linear dilaton solution of string theory,
which we consider in \S3.
We study in particular the type IIA and type IIB Neveu-Schwarz
fivebrane solutions (for which $d=6$).  In this case, the
string theory solution again has a diverging $F^2$ term while
the Einstein term survives with a finite coefficient upon dimensional
reduction from $7d$ to $6d$.  The gauge field propagates as
if in $7d$ flat space in the linear dilaton solution.
However, the IIA solution gets corrected to one
localized along the eleventh dimension of M-theory \itzhaki, so that
the symmetry is effectively Higgsed.  In the IIB case one also
finds a lifting of the RR two-form potential from effects
occurring in a region where the linear dilaton solution
has broken down.

In \S4\ we discuss some aspects of the physics of charged black
holes that make the $d+1$ and $d$ dimensional descriptions of
these systems consistent.
Finally in \S5\ we discuss other long-wavelength solutions which
naively exhibit this effect and discuss prospects for realizing
them in M theory.

One possible application is to the problem of
compactifying matrix theory down to four dimensions.  This
requires considering D0-branes in a IIA compactification
down to three dimensions.  This  is problematic in part
because of the infinite classical electrostatic self-energy of the
D0-brane in $3d$.  If the electric field lives in $4d$, while
gravity is localized to $3d$, this problem may be avoided.

\newsec{Cutoff AdS Space}

In AdS spaces cut off by a  Poincare-invariant
``Planck brane'', the metric can
be written
\eqn\adsmet{ds^2= e^{-2|r|/L}dx_{||}^2+dr^2}
where $x_{||}$ refers to the dimensions along the brane.
The dimensional reduction of the Einstein term in the
$5d$ action gives a finite $4d$ Planck scale $M_4$ \RS:
\eqn\planck{M_4^2\propto M_5^3\int_0^\infty dr e^{-2r/L}}
where $M_5$ is the $5d$ Planck scale.
On the other hand, if we introduce a Maxwell field, its
kinetic term in $4d$ is divergent.  In terms of an infrared cutoff
$R$, the corresponding integral is
\eqn\gauge{{1\over e^2}\sim \int_0^R dr \propto R,}
which diverges as $R\to\infty$.  From the metric \adsmet,
this cutoff scale $R$ on the coordinate $r$ corresponds
to -$L$log ($k_0 L$) where $k_0 \ll L^{-1}$ is an infrared momentum cutoff
along the four dimensions parameterized by $x_{||}$.
So this effect is quite conventional in four dimensions:
the charge is screened by the charged matter in the $4d$
conformal field theory dual to this background
\adscft.\foot{In the realization of $AdS_5$ in the
IIB compactification on $S^5$, there is a factor
of $N^2$ in the expression for the renormalized $1/e^2$.
This comes from the factor $Vol(S^5)/(l_s^8g_s^2)=N^2/L^3$
in front of the gauge field kinetic term.}

This result agrees with that obtained by a direct calculation
of the electrostatic potential of a charge localized on the Planck
brane.  (See \gaugebulk\ for similar calculations for
gauge fields in \RSII, and for example \duff\gub\RSBH\
for analogous calculations
of corrections to the gravitational propagator.)
The electrostatic potential is found by integrating the
Maxwell's equation
\eqn\maxwell{\nabla_M F^{MN} = - Q J^N}
in the background geometry \adsmet\ in the electrostatic
approximation, when the vector potential is $A_M = (\Phi, 0, 0, 0, 0)$
and the density current is $J^M = {1 \over
\sqrt{g}}\delta^{(4)}(x-x_0)(1, 0, 0, 0, 0)$ (here $x$ stands for all
spatial coordinates in \adsmet).
To compute the potential, it is convenient to choose the
coordinates such that the metric \adsmet\ is conformally flat.
Defining $|z|+L = L \exp(|r|/L)$, the metric becomes
\eqn\adsmetconf{ds^2= {L^2 \over (|z|+L)^2} (dx_{||}^2+dz^2),}
while the Maxwell's equations \maxwell\
reduce to
\eqn\maxexp{(|z|+L) ({\Phi' \over |z|+L})' + \vec \nabla^2 \Phi =
Q \delta^{(3)}(\vec x - \vec x_0) \delta(z).}
Since the cutoff $AdS_{5}$ space is realized with the orbifold
symmetry $r \rightarrow -r$, enforcing this symmetry requires
$\Phi(z, \vec x) = \Phi(|z|, \vec x)$, and so defining the variable
$\rho = 1+|z|/L$ and
Fourier transforming in the longitudinal
spatial directions, \maxexp\  becomes
\eqn\fourmax{\rho^2 {d^2 \tilde\Phi \over d\rho^2} - 
\rho {d \tilde\Phi \over d\rho}
- \rho^2 \vec k^2 L^2 \tilde\Phi 
= ({Q L} - 2{d \tilde\Phi \over d\rho}) \delta(\rho - 1).}
It is straightforward to see that the only solutions of the homogeneous part of
this equation which are regular on the $AdS$ horizon $|z|\rightarrow \infty$
are $\tilde \Phi = A \rho K_1( k L\rho)$,
where $K_n(x)$ are the Macdonald functions 
of index $n$ (also known as modified Bessel functions
of the third kind) and $k=\sqrt{\vec k^2}$.
The potential is then determined by choosing the integration
constant $A$ to satisfy the boundary condition
$ 2{d \tilde\Phi \over d\rho}\biggl|_{\rho=1}=QL$,
as required by the $\delta$-function source in \fourmax.
The solution is
\eqn\grf{\tilde\Phi(\rho,\vec k) 
= - {Q \rho \over 2 k K_0( k L)} K_1(k L\rho).}
Returning to the original coordinates of the cutoff $AdS$ space \adsmet\
and Fourier transforming
back, we find the electrostatic potential of a particle
located on the cutoff
brane:
\eqn\poten{\Phi(r,\vec x) = - {Q\over 2} e^{|r|/L} \int {d^3\vec k \over (2\pi)^3}
{K_1( k L e^{|r|/L}) \over k K_0(kL)} e^{i \vec k \cdot (\vec x - \vec x_0) }.}
To diagnose screening, we consider the potential at very
long distances $|\vec x - \vec x_0| \gg L$ along the cutoff brane $r=0$. This
is dominated by the $k \rightarrow 0$ contributions to the integral, where the
momentum space potential is
\eqn\limit{\tilde\Phi(0, \vec k) \rightarrow {Q \over 2 k^2 L \ln(kL/2)},}
which is precisely the potential of a screened
charge $Q(k) = {Q \over 2 L \ln(kL/2)}$,
confirming the expectation.
We note that in the case of gravity
localized to an intersection of $n$ $2+n$-branes in
$AdS_{3+n+1}$ \addk\ a similar analysis shows that the screening
effect persists.


In the case of $AdS_{d+1}$ for $d>4$, the integral determining the
gauge coupling is finite.  In the case $d=3$, one also obtains
a result consistent with
screening behavior:  the electrostatic potential goes like
$1/x_{||}$ along the brane.  This case might be of interest
for a matrix theory formulation of $4d$ gravity.

It is instructive to consider the analogous calculations for
higher-form gauge potentials.  The kinetic energy is (for
a gauge field with a $q$-form field strength)
\eqn\qkin{\int d^dx dr \sqrt{g} F_{\mu_1\dots\mu_q}g^{\mu_1\nu_1}\dots
g^{\mu_q\nu_q}F_{\nu_1\dots\nu_q}.}
Integrating this over $r$ up to a cutoff $R$ yields an effective
charge
\eqn\qcharge{\eqalign{{1\over e_q^2}\propto ~~~~~~
&e^{{R\over L}(2q-d)}\sim {1\over{k_0^{2q-d}}}~~~~2q\ne d\cr
&R\sim log(k_0)~~~~2q=d\cr}}
where $k_0$ is an infrared momentum cutoff along the
$x_{||}$ directions.
So for $2q\ge d$, the charge is effectively screened at long distance,
more strongly for higher-form field strengths.
It would be interesting to understand microscopically
how this screening occurs--perhaps it arises from spherical $q-2$-branes, which
can develop multipole moments as discussed by Myers in
\rcm .
One caveat is
that additional interactions can cause the
$q-1$-form potential to be lifted
at low energies instead.
In the next section we will see examples of this possibility.

It is interesting that the effects of light charged matter arise
from the AdS part of the gravity solution alone.  In
known supersymmetric M-theoretic
realizations of AdS solutions, there is a five-dimensional Einstein
manifold whose isometries yield gauge symmetries and whose Kaluza-Klein
excitations provide charged matter.  At the level of low-energy
field theory, one could contemplate a situation where the
gauge field was present in the bulk but charged matter lived only
on the brane.  In such a situation, the $4d$ behavior of
the Maxwell field would be as if there were screening by
light charges, but there would be no dynamical charges in
the bulk that could be excited.\foot{Something somewhat analogous
happens in at the conifold singularity in Calabi-Yau
moduli space if the string coupling is taken to zero before
the singularity is reached:  then the light wrapped D-branes
that are usually responsible for the screening of the
RR charge in that system \conifold\ are decoupled, and the effect comes
from the singular ``throat CFT'' \phases.}  It seems likely
that such a situation does not occur in M theory realizations
of $AdS$  (this is certainly true in the case of
the supersymmetric realizations that are most familiar).
In particular, as discussed in
\S4, bulk charged matter plays a crucial role in black hole physics
in these systems.

\newsec{$6d$ Little String Theories}

Consider the $N$-NS 5-brane solution of type II string 
theory \fivebrane.  It
has a string-frame metric and dilaton
\eqn\lindilone{\eqalign{
&ds^2=dx_6^2+dr^2+l_s^2Nd\Omega_3^2\cr
&\phi=\alpha r\cr
}}
with $\alpha=1/l_s\sqrt{N}$.
There is a flux of the three-form NS field strength $H$ which
stabilizes the $S^3$ component of the geometry.
We want to consider the behavior of gravitons and of RR gauge
potentials in this background.

The decoupled
throat theory has this metric with $r\in (-\infty, +\infty)$.
For $N\le 16$ we can cut it off by considering it as part of
a compactification, so the full metric is rather complicated
but asymptotes to \lindilone\ down the throat
$r\to +\infty$ of the NS5's.
This is similar to the proposal \hermann\ for realizing
the Randall-Sundrum background in string theory via compactification.
The compactification which does this most simply is on
the moduli space of type II on $T^4/I_4(-1)^{F_L}$ as studied
by Kutasov \kut\ and by Sen \sen.
Note that in the string-realized
$AdS_{d+1}$ cases this was not a possibility, since a compactification
transverse to the brane would explicitly break the SO(10-d) symmetry
of the $S^{9-d}$ surrounding the brane.
Here we are not making
use of the analogue of that symmetry, but instead are using the
U(1) generated by the IIA RR 1-form potential.
To consider $N>16$, we would need a ``Planck brane'' of the
sort considered in \RS\ in order to cut off the solution
and bind gravity.  We do not know if this has a precise
realization in M theory, but will assume so in discussing
this case.

The string coupling grows down the throat of the solution,
so most calculations are out of control far down the throat.
The corrections to the solution as discussed
for example in \itzhaki\
will be important.

The string-frame ten-dimensional action is
\eqn\strac{\int d^6x dr d\Omega_3 \biggl[
e^{-2\phi}(R+(\partial\phi)^2)+K_{RR}^2\biggr]}
where $K$ is the field strength for the RR U(1) gauge field of
type IIA string theory, or the field strength for the 2-form
RR gauge potential of type IIB string theory.

Consider the perturbation $g_{\mu\nu}\to g_{\mu\nu}+h_{\mu\nu}$
in this linear dilaton background,
where $\mu,\nu$ run along the 6 dimensions of the 5-branes.
Let us dimensionally reduce the system to $5+1$ dimensions,
and write an effective action for the graviton $h_{\mu\nu}$.
Though we here work in string frame, the Einstein frame description
of course yields the same results.

In determining the 6d Einstein term we must integrate over
the extra 4 dimensions in the NS5 background.  The contribution
of the rest of the compactification is finite, so the only
issue is the throat.
The integral over $r$ relevant for the $6d$ Planck scale goes like
\eqn\rint{M_6^4\sim m_s^5\int dr e^{-2\alpha r}={m_s^5\over{2\alpha}}}
which is finite in terms of the string mass scale
$m_s=1/l_s$.  (Here for simplicity we have placed the
Planck brane at $r=0$.)

Now do the same for the RR gauge field.  It does not couple to
the dilaton in string frame (in its realization with a standard
gauge invariance, rather than one intertwined with the dilaton)
\eddiverse.
So the integral over $r$ is infinite.  This already shows a
definite difference from a standard gauge symmetry:  as reviewed
above, in the AdS cases this integral is finite
for 1-form gauge potentials in $d>4$,
reflecting the fact that the higher-dimensional QFTs do not screen
the charge via quantum effects.

Furthermore, Maxwell's equations
are easy to solve here, since we are in flat space with no
extra direct couplings to $\phi$ as far as
the RR gauge field is concerned.  The power law falloff
of the electric field will
be as in 7 dimensions rather than 6.  So it naively looks like this
is a case where the gauge field propagates in one higher dimension
than gravity, which is ``localized'' in 6d.
We will find, however,
that this is not true;
the symmetry is Higgsed.

In \stringglob, arguments against global symmetries in
perturbative string theories were presented.
It should be noted that the effect we see here at the level
of perturbative string theory is not in contradiction with
the results of \stringglob\ for two reasons:
(i) perturbation theory breaks down down the throat and (ii) the
throat is noncompact (there is a continuum of modes).

\subsec{Type IIA}

In the type IIA string theory, the strong coupling limit occuring
down the throat of the solution is better described by the
eleven-dimensional limit of M-theory.  In this description, the
eleventh circle $S^1_{11}$ is expanding down the throat of the solution.
Lifting the solution \lindilone\ directly to eleven-dimensional
supergravity gives a configuration where M5-branes are ``smeared''
over the eleventh dimension.  For a finite number $N$ of NS5-branes,
this solution is microscopically corrected to one in which
$N$ M5-branes are localized at points on $S^1_{11}$ \itzhaki.
For an infinite number of 5-branes, one can contemplate the smeared
solution corresponding to a continuus placement of the infinite
number of branes along the circle, as in analogous cases
studied in the context of the AdS/CFT correspondence \smeared.
However, as we will review,
the scaling of parameters that leads to this solution
invalidates the supergravity approximation and more importantly
removes the localization of the graviton.

\noindent{\it Microscopics}

Starting with finite $N$, we can find a microscopic embedding of the cutoff
solution into M theory.  This is obtained by the construction of \kut,
in which the NS5-branes sit at points in a compactification manifold.
The full solution is \itzhaki\oferetal\
\eqn\elevmet{\eqalign{
&ds^2=l_p^2f^{-{1\over 3}}\biggl[dx_{||}^2+f(dy_{11}^2+dU^2
+U^2d\Omega_3^2)\biggr]\cr
&f=\sum_{n=-\infty}^{\infty}\sum_{i=1}^N{1\over
{[U^2+(y_{11}-y_i+{n\over l_s^2})]^{3\over 2}}}\cr
}}
where the $y_i, i=1,\dots, N$ denote the positions of the $N$ branes
on $S^1_{11}$.  The metric is written here in terms of
the coordinate
$U={\sqrt{N}\over l_s^2}e^{{r\over{\sqrt{N}l_s}}}$.

Since this solution breaks the translation symmetry
along $S^1_{11}$, the U(1) RR 1-form potential is Higgsed.  This
is an infrared effect in the $6d$ description.  The microphysics of
M theory therefore avoids the issue of a global symmetry arising
in $6d$, by substituting spontaneous breaking of the symmetry.

This Higgsing persists in the appropriate $N\to\infty$ limit.
In terms of these coordinates (which correspond to canonically
normalized VEVs of fields in the (2,0) conformal field theory \oferetal)
the periodicity of $y_{11}$ is $1/l_s^2$.  From \lindilone\
(with $\alpha={1\over{l_s\sqrt{N}}}$) it is clear that to localize
gravity as $N\to\infty$ we need $l_s\to 0$ so that
$l_s\sqrt{N}$ does not diverge in the limit.  This is also
required for having a valid supergravity approximation everywhere
\minsei.  For evenly-spaced branes, the spacing between branes is fixed:
\eqn\spacing{\Delta y=y_i-y_{i-1}={1\over{l_s^2N}}.}
So what happens in this limit is that the eleventh circle $S^1_{11}$
expands as $N\to\infty$, leaving the spacing between branes
fixed and the symmetry of interest here broken.  The smeared solution,
which preserves the symmetry, could only pertain to a different
scaling which removes the localization of gravity.

\subsec{Type IIB}

In the type IIB theory, the relevant gauge potential is the
RR 2-form $B_{RR}$.  In this case, as one increases $r$,
the solution
\lindilone\
crosses over to the D5-brane solution and then to
a description in terms of the SYM theory formed by the light
open strings living on the D5-brane \itzhaki.  In the
D5-brane and SYM descriptions, the original RR $B$ field
becomes an NS $B$ field.

In the open string description, the $B$ field couples to the
worldvolume $U(1)$ gauge field via the
Stuckelberg coupling \wittbound
\eqn\coupling{
S=\int d^{10}x |dB|^2+\int d^6x (B-F)^2}
The second term effectively gives a mass to the $B$ field.
Thus in this case also, the massless gauge boson gets lifted down
the throat of the solution instead of leading to a global
symmetry in $6d$.  This is consistent with the T-duality to the IIA
case on a circle.

\newsec{Black Hole Physics}

Even with the conventional $d$-dimensional understanding
that we have come to of the physics
of $d+1$-dimensional gauge fields in our systems, the
$d+1$-dimensional picture raises interesting questions
about black hole physics.  We will here provide a
qualitative discussion of some of these issues; it
would be interesting to find concrete black hole solutions
in these backgrounds to study.
Schwarzschild black holes
were considered in \RSBH; in the systems we are
considering here the generalization to
charged black holes is of interest.

In the $AdS$ cases, the symmetry is unbroken and the
charge is conserved (and screened in low
enough dimension).  Suppose there is an extreme or nearly
extreme black hole of charge $Q$ and mass
$M$ centered on
the Planck brane in $AdS_5$.  Its charge can be measured by Gauss' law
in five dimensions, and this charge is conserved overall in
the system.  In the four-dimensional description, the charge
is screened, and could not be measured at long distance.
One expects
quantum mechanically a charged black hole to be quickly neutralized
at long distances by the light charged conformal
field theory matter in the
system (in analogy to the familiar effect in ordinary
QED in black hole backgrounds \gibbons).

How does this occur from the five-dimensional point of view?
In order to see the neutralization from
this point of view, we need the electric field to be strong enough to make
it energetically favorable for charged matter to be
pair-produced and to draw the negatively charged member
of the pair to the black hole (against the gravitational attraction
toward the $AdS$ horizon).
Let us assume we have charged matter of mass $m$ and
charge $q$ in the
bulk theory.  From \poten\ for large $|z|$ one finds
\eqn\electfld{
F^{0z}~~~\propto ~~~ {{Q |z|^2}\over L^5}}
So the strength of the electric field grows toward the $AdS$
horizon.  Since this will eventually dominate over the mass,
we expect an analogue of the Schwinger calculation to
imply pair production (though we have not calculated this effect
in our background directly).  In terms of the forces on the produced
pair, the source $Q$ leads to an electromagnetic force on the
particle of charge $q$ in the
$z$ direction of magnitude
\eqn\emforce{f^z_{E.M.}~~\propto ~~{{qQ z}\over L^4}.}
Here we took the particle to be at rest and evaluated the relation
$f^\mu_{E.M.}=qF^\mu{}_\nu {{dx^\nu}\over{d\tau}}$ in our background
field configuration \adsmetconf\electfld.  There is also a gravitational
attraction to the black hole, as well as a gravitational
attraction toward the $AdS$ horizon.  The latter effect goes like
\eqn\AdSforce{f^z_{AdS}~~\propto ~~ -m\Gamma^z_{\mu\nu}{{dx^\nu}\over{d\tau}}
{{dx^\mu}\over{d\tau}}\sim -m z {1\over L^2}}
{}From \emforce\ and \AdSforce\ it is clear that for large
enough $Q$, the electromagnetic force will dominate.
We therefore expect dielectric breakdown from pair production of
the charged matter of mass $m$ to
become possible
at large enough $|z|$.
For small $Q$ (and small size relative
to the AdS radius),  the object does not
constitute a black hole from the $4d$ point of view in any case.

Another effect to consider is
the quantum stability of the
localization of
the charge $Q$ on the brane.  Particles in
the bulk fall toward the $AdS$ horizon, and therefore
it seems clear that the charged source
will ultimately tunnel into the bulk and fall down
the throat.  In the absence of a genuine
embedding of the Planck brane
into M theory, we cannot calculate the rate for this
tunneling process.  To a $4d$ observer it would look like the conserved
charge is spreading out.  The boundary of this region of
spreading charge may behave like a membrane, and
the $5d$ gravity description might provide a way to study
membrane nucleation processes.

One can similarly consider charged objects in the IIA system
of \S3\ in the regime where the ``throat'' solution \lindilone\
applies.
The D0-branes in this background (which are momentum modes around
$S^1_{11}$) become light down the throat, exponentially in $r$:
\eqn\Dzeromass{M_{D0}={1\over{e^\phi l_s}}
\sim {1\over l_s}e^{-2\alpha r}}
On the other hand, the electric field emanating from a
source of charge $Q>0$ at the Planck brane decays like a power:
\eqn\electricfld{\vec E\sim {Q\over{(r^2+x_{||}^2)^{5\over 2}}}.}
We again expect that for large enough $r$, the light D0-branes
will be pair-produced and neutralize a charged black hole in
this system as well.


There is another intriguing aspect to the physics of charged
objects in this sort of system.
In $d+1$ dimensions one can measure the charge classically
using Gauss' law.  The charge that one measures this
way is the bare unscreened charge $Q$.
This procedure must translate into some operation
in the $d$-dimensional description of the system.
According to the IR/UV relationship in holography  \susswitt, the
$d+1$-dimensional
Gauss' law measurement always uses information that is longer-wavelength
than the size of the object in the $d$-dimensional description.

\newsec{More General Solutions}

We have seen that in cases where there is a brane interpretation
of a warped metric, M theory conspires to prevent a truly
higher-dimensional gauge field from arising in a $d$-dimensional
gravity theory.  We expect it is likely that this happens rather
generically.  Still, it is interesting to consider backgrounds
which, like those discussed here, have such an effect naively
in a long-wavelength analysis---but
whose microscopic behavior is not yet understood.

The linear dilaton solution \lindilone\ arises much more generally
than in the NS5-brane solutions (in general with
the $d\Omega_3^2$ piece replaced with something else).
Geometrical singularities such as the conifold singularity
routinely resolve into a throat with linear dilaton behavior.
This can be seen from the description of such compactifications
using the techniques of \phases.  Some of these cases in fact
descend from those considered above via K3 fibration.  It would
be very interesting to systematically analyze the behavior
of gauge bosons in many other classically singular geometries.

Another place where linear dilaton solutions arise is in
noncritical perturbative
string backgrounds with a tree-level cosmological term.
There the linear dilaton solution \lindilone\ exists with
$\alpha^2\propto (D-D_{crit})/l_s^2$.  In this
case, as before, a naive calculation would suggest the
potential for a delocalized gauge boson.  Since the dilaton
grows along the direction $r$ in the solution, strong
coupling arises and corrections will be important.
With current technology we cannot say whether this will
always lead to conventional $4d$ behavior or whether
it is conceivable that sometimes
$4d$ effective field theory will break down in such a way
that a global symmetry of the kind we have been contemplating
can persist.

One could similarly consider the dilaton gravity solutions
of the type considered in for example \singsolutions.  As we discussed
in the introduction, the coupling of the scalar to
the $F^2$ term can (quite generically) be such that
again a naive calculation of the coupling would suggest
a globalization of the symmetry in $4d$.  However, as
above, corrections will be important (and will rule out
some subset of these solutions altogether).

The gravity backgrounds we considered in the bulk
of the paper have known holographic
descriptions in terms of a brane worldvolume
theory which reduces to ordinary
$d$-dimensional effective
quantum field theory at low
energy.  Most backgrounds have no known holographic
description (for some relatively recent
considerations of the general features required
see for example \tomofer), and it is not yet clear
what precise form holography will take in a generic background.
It will be interesting to study the structure of gauge
symmetries in various dimensions once more general backgrounds
are understood.

In any case, as it stands, in this paper we have gathered
further evidence for the robustness of the arguments
against global symmetries in the context of gravity.

\centerline{\bf Acknowledgements}

We would like to thank T. Banks, S. Dimopoulos, M. Duff,
S. Kachru, A. Lawrence, J. Maldacena, L. Randall, M. Schulz, S. Shenker, 
and especially E. Witten,
for very useful discussions.  We understand N. Arkani-Hamed, M. Porrati,
and L. Randall have results overlapping with ours (particularly
in section 2).   The work of E.S. is supported by a
DOE OJI grant, by the A.P. Sloan
Foundation, and
by the DOE under contract DE-AC03-76SF00515.
The work of N.K. and  L. S. is supported in part by NSF grant
980115.

\listrefs
\end